\documentstyle[titlepage,12pt]{article}
\def\beq{\begin{equation}}
\def\eeq{\end{equation}}
\def\barr{\begin{eqnarray}}
\def\earr{\end{eqnarray}}
\def\ket#1{\left |#1\right\rangle}
\def\bra#1{\left\langle#1\right |}
\begin{document}
\begin{titlepage}
\renewcommand{\thefootnote}{\fnsymbol{footnote}}
\vskip 1.0cm
\begin{center}
{\bf Quantum Hall effect from soliton equation}
\end{center}
\vskip 0.5cm
\begin{center}
Hideaki Hiro-Oka\footnote[1]{Fellow of Japan Society for the Promotion of
                             Science}%
                \footnote[2]{e-mail: hiro-oka@phys.metro-u.ac.jp} and %
        Satoru Saito\footnote[3]{e-mail: saito@phys.metro-u.ac.jp}
\end{center}
\vskip 1.0cm
\begin{center}
{\it Department of Physics\\
Tokyo Metropolitan University\\
1-1 Minami-Ohsawa Hachioji, Tokyo 192-03, Japan}
\end{center}
\vskip 3cm
\begin{flushleft}
{\bf Abstract}
\end{flushleft}

The Laughlin function of quantum Hall effect is shown to satisfy Hirota's
bilinear difference equation with certain coefficients a little different
from the
KP hierarchy. Vertex operators which constitute blocks of solutions
generate a B\"acklund transformation. Besides the Laughlin function,
the equation admits soliton solutions.

\begin{flushleft}
\end{flushleft}
\end{titlepage}


The experimental discovery \cite{Klitzing}\cite{Tsui} of the integer and of
the fractional quantum Hall effect (QHE) \cite{Prange} provided us a
fascinating object of study in various
fields. For example, due to the exact quantization of plateaux this
phenomenon is utilized
to determine the fine structure constant $\alpha$ precisely. Many authors
have tried to
clarify the mechanism of the QHE which can never be understood at the semi-%
classical level.
The picture of incompressible quantum fluid by Laughlin \cite{Laughlin} is
the basis in
understanding the existence of plateaux in the Hall conductivity.
As the excitation the quasi-hole or the quasi-particle is yielded. The
remarkable thing is that they obey the fractional statistics. They are
expressed as composite particles with some flux quanta in the Chern-%
Simons gauge theory \cite{Wilczek}.
Their collective motion is analyzed by the mean field theory with
random phase approximation, collective field theory \cite{Hiro-Oka}, and so on.
Then, this composite object picture plays an important role in the hierarchy
problem by Jain \cite{Jain}, which is different interpretation from that by
Haldane \cite{Haldane} and Halperin \cite{Halperin}.
Fubini pointed out the similarity between Laughlin's wave
function and the Veneziano amplitude introduced in the theory of particle
physics \cite{Fubini1}\cite{Fubini2}.
He described the wave function in terms of the vertex operator of particle-%
string interaction. Together
with the work by Moore and Read \cite{Moore} it became the starting point of
investigating
the QHE by using the conformal field theory.

In this article we shall try to shed new light to Laughlin's theory from
the point of view of the soliton theory. Since the correspondence between the
soliton theory and the string theory has been established
\cite{Saito}\cite{Sogo} our observation
will complete the link of three different objects in physics, the QHE, soliton
theory, and string theory. To this end we first derive vertex operator which
generates B\"acklund transformation of Laughlin's wave functions and show
the correspondence of this operator with the one of soliton theory. Secondly we
derive a bilinear difference equation of Hirota type \cite{Hirota} which
relates Laughlin's
wave functions with different values of filling factor. Once we establish
the equation characterizing Laughlin's wave function it enables us to find
other solutions and investigate their physical meaning. In fact we will find
a soliton type of solution besides Laughlin's wave function.


Let us start with reviewing briefly the notion of Laughlin state of the
quantum Hall phenomena \cite{Prange}. In quantum mechanics the electron
system in two
spatial dimension with the
magnetic field perpendicular to the plane is exactly solved. The Hamiltonian
is given by
\beq
H={1\over {2m}}\left(-i\nabla-{e\over c}A\right)^2\label{hamiltonian},
\eeq
with the symmetric gauge $A={1\over 2}B\times r$.
The electron semi-classically undergoes a
cyclotron motion with the frequency $\omega_c=eB/mc$. This system is
characterized by the magnetic length $l=\sqrt{2\hbar c/eB}$ which is the
cyclotron radius such
that one flux quantum exists in the encircled area. In what follows we use
the natural unit $\hbar=c=1$ and set $l=m=1$ for simplicity.

By using the following operators
\beq
a={1\over 2}z+{\bar\partial},\quad a^\dagger={1\over 2}{\bar z}-\partial
\label{operator1}
\eeq
\beq
b={1\over 2}{\bar z}+\partial,\quad b^\dagger={1\over 2}z-{\bar\partial}
\label{operator2}
\eeq
which satisfy the commutation relation $[a,a^\dagger]=[b,b^\dagger]=1$,
the Hamiltonian and the angular momentum are rewritten as
\beq
H=2a^\dagger a+1\label{hamiltonian2}
\eeq
and
\beq
J=b^\dagger b-a^\dagger a,\label{angular}
\eeq
respectively where $z=x+iy$ and $\partial=\partial/\partial z$,
${\bar \partial}=\partial/\partial {\bar z}$. The energy is discretized
(Landau level). As can be seen,
this system is highly degenerate
on the angular momentum. The ground state is determined by the condition
\beq
a\psi_{00}=b\psi_{00}=0,\label{gcond}
\eeq
e.g.,
\beq
\left({1\over 2}z+{\bar\partial}\right)\psi_{00}=0\label{geq}
\eeq
and the wave function is given by
\beq
\psi_{00}={1\over{\sqrt{\pi}}}e^{-{1\over 2}|z|^2}.\label{qwf1}
\eeq
We obtain the generic form of energy $n$ and angular momentum $m$ by acting
the creation operators:
\beq
\psi_{nm}={(b^\dagger)^{m+n}\over{\sqrt{(m+n)!}}}%
{(a^\dagger)^n\over{\sqrt{n!}}}\psi_{00}\label{wf}
\eeq
In the case of strong magnetic field only the lowest Landau level is realized
due to the large energy gap.
We concentrate our attention to the lowest Landau level wave function such as
\beq
\psi_{0m}={z^m\over{\sqrt{m!\pi}}}e^{-{1\over 2}|z|^2}.\label{gwf2}
\eeq
In many body system with no interaction between electrons one can obtain the
wave function in the tensor products of the Hilbert space of one body wave
function by taking into account the statistics.
The quantum Hall system, however, has not only the Coulomb interaction
between electrons but also the impurity effect.
It is, therefore, convenient to consider the variational function such as
Laughlin's wave function since it is hard to obtain the exact many body
wave function.

{}From the analysis of two- and three-body problems Laughlin introduced the
following variational function
\beq
\psi(z_1,z_2,\ldots,z_N)=\prod_{i>j}(z_i-z_j)^me^{-{1\over2}\sum|z_i|^2}.
\label{laughlin1}
\eeq
Here $m$ is an odd number due to Pauli's principle. According to the one
component plasma analogy $m$ is given by the inverse of the filling factor
$\nu$.

Let us ignore, for a moment, the gaussian factor and consider the Jastrow part,
\beq
f(z_1,z_2,\ldots,z_N)=\prod_{i>j}(z_i-z_j)^m.\label{jastrow}
\eeq
This satisfies the following equation
\beq
\left(\partial_i-m\sum_{i>j}{1\over{z_i-z_j}}\right)f=0,\label{kz}
\eeq
which is called Knizhnik-Zamolodchikov equation for $c=1$ conformal field
theory. From this fact it is natural to express (\ref{jastrow}) as a
correlation function of vertex operators of the conformal field theory.

The vertex operator was first introduced in the context of study of the dual
resonance model. This was applied to the Kadomtsev-Petviashvili (KP)
hierarchy in ref. \cite{Kashiwara}\nocite{Date1}-\cite{Date2} as we shall
mention below.
The vertex operator $V(k,z)$ for the Veneziano amplitude is expressed in
terms of the string coordinate $X(z)=X^{+}(z)+X^{-}(z)$
\beq
V(k,z)=e^{ikX^{+}(z)}e^{ikX^{-}(z)},\label{vo1}
\eeq
where
\barr
X^{+}(z)&=&q-\sum_{n>0}{i\over\sqrt{n}}a_{-n}z^n\label{scord1}\\
X^{-}(z)&=&-ip\ln z+\sum_{n>0}{i\over\sqrt{n}}a_n z^{-n}\label{scord2}
\earr
and $k$ is the momentum of the external line. The coefficients of $z$
satisfy the following commutation relations,
\beq
[a_n,a_{-n'}]=\delta_{nn'},\quad [q,p]=i.\label{comrel}
\eeq
The Veneziano amplitude is described as an $N$ point correlation function,
{\it i.e.}
\beq
\bra{0} V(k_1,z_1)V(k_2,z_2)\cdots V(k_N,z_N)\ket{0}%
=\prod_{i>j}(z_i-z_j)^{k_ik_j}.\label{veneziano1}
\eeq
provided the vacuum state is such that
$a_n\ket{0}=p\ket{0}=0$. Comparing the Veneziano
amplitude (\ref{veneziano1}) with the Jastrow part (\ref{jastrow}) we find
out that $f$ is described by
the vertex operators (\ref{vo1}) in the case of constant momentum
$k_i^2=m$,~$\forall_i$.
The point is that $\sum k_i=0$ due to the momentum conservation. Thus we
must consider $\bra{-K}$ as a screening charge, where $\sum k_i=-K$,
in advance. See ref.\cite{Fubini1} for more discussions.

Now we notice that another realization of (\ref{vo1}) is possible if we write
\beq
\left.
\begin{array}{ll}
a_{-n}&=\sqrt{n}t_n\\
a_n&=\displaystyle{{1\over\sqrt{n}}{\partial\over{\partial t_n}}}
\end{array}
\right\}\quad n\geq 1,\label{soperator1}
\eeq
\beq
q=-it_0,\quad p={\partial\over{\partial t_0}},\label{soperator2}
\eeq
\barr
\tilde{V}(k,z)&=&\exp\left(-{1\over 2}|z|^2+k\sum_{n=0}t_nz^n\right)%
           \nonumber\\
           &&\times\exp\left[k\left(\ln z{\partial\over {\partial t_0}}%
           -\sum_{n>0}{1\over n}{\partial\over{\partial t_n}}z^{-n}%
           \right)\right].\label{vo2}
\earr
Since the gaussian factor is not affected by the operation of $a_n$ we
include it in the definition of the vertex operator.
The product of two operators is of the form
\begin{eqnarray}
\tilde{V}(k_1,z_1)\tilde{V}(k_2,z_2)&=&(z_1-z_2)^{k_1k_2}\exp\sum_{j=1,2}%
\left(-{1\over 2}|z_j|^2+\sum_nt_nk_jz_j^n\right)\nonumber\\%
&&\times\exp\sum_{j=1,2}\left(k_j\ln z_j{\partial\over{\partial t_0}}
-\sum_{n>0}{1\over n}k_jz_j^{-n}{\partial\over {\partial t_n}}\right),
\label{vot}
\end{eqnarray}
which enables us to write Laughlin's wave function as,
\beq
\left.\tilde{V}(k_1,z_1)\tilde{V}(k_2,z_2)\cdots \tilde{V}(k_N,z_N)\cdot 1%
\right|_{t_n\rightarrow 0}%
=\prod_{i>j}(z_i-z_j)^{k_ik_j}\exp\left(-{1\over 2}\sum_i|z_i|^2\right),
\label{laughlin2}
\eeq
with $k_i^2=m$, $\forall_i$.
It turns out that this operator is quite similar to the vertex operator for
KP hierarchy up to the gaussian factor and zero mode of the string coordinate
(\ref{scord1}) (\ref{scord2}).
In fact, if $k_i$'s are the same sign in (\ref{vo2})
we obtain
\beq
\tilde{V}^2(k_j,z_j)=0,\label{vo3}
\eeq
moreover,
\beq
\exp \tilde{V}(k_j,z_j)=1+\tilde{V}(k_j,z_j).\label{vo4}
\eeq
It is nothing but this fermionic nature which provides solitons as solutions
to the KP hierarchy. The vertex operator for the KP hierarchy is given by
\barr
V_{\rm KP}(z,w)&=&\exp\left(\sum_{n>0}(z^n-w^n)t_n\right)%
            \exp\left[-\sum_{n>0}{1\over n}\left({1\over z^n}%
                                   -{1\over w^n}\right)%
                      {\partial\over {\partial t_n}}\right]\label{vokp}\\
&\propto&(z-w)\tilde{V}(1,z)\tilde{V}(-1,w).
\earr
This is the operator which generates the B\"acklund transformation of the KP
hierarchy.
This fact suggests that the quantum Hall system is somehow related to
the soliton
theory. In other words, it might have a new interpretation
of the QHE by means of the language in the soliton theory.

It has been proved that the
soliton equations in the KP hierarchy are governed by HBDE \cite{Hirota}.
The Laughlin functions are, however, different from the $\tau$-function of the
KP hierarchy. Therefore an interesting question is whether exist coefficients
of HBDE which admit Laughlin's functions as solutions.

The general form of HBDE is given by
\beq
F(D_1,D_2,D_3)f\cdot f=\left(\alpha e^{D_1}+\beta e^{D_2}+\gamma e^{D_3}%
\right)f\cdot f=0,\label{hbde1}
\eeq
under the constraint
\beq
F(0,0,0)=0.\label{cond1}
\eeq
$D_i$ is the Hirota derivative with
respect to the variable. Writing this explicitly we have
\barr
 \alpha f(\mu+1,\nu,\lambda)f(\mu-1,\nu,\lambda)&&\nonumber\\
+\beta  f(\mu,\nu+1,\lambda)f(\mu,\nu-1,\lambda)&&\nonumber\\
+\gamma f(\mu,\nu,\lambda+1)f(\mu,\nu,\lambda-1)&=&0.\label{hbde2}
\earr
It is, however, more convenient to consider it in other set of variables:
\barr
 \alpha \tilde f(k_1+1,k_2,k_3)\tilde f(k_1,k_2+1,k_3+1)&&\nonumber\\
+\beta  \tilde f(k_1,k_2+1,k_3)\tilde f(k_1+1,k_2,k_3+1)&&\nonumber\\
+\gamma \tilde f(k_1,k_2,k_3+1)\tilde f(k_1+1,k_2+1,k_3)&=&0. \label{hbde3}
\earr
This can be obtained by the use of the following sequential variable
transformations:
\beq
2\mu=K_2+K_3,\quad2\nu=K_3+K_1,\quad 2\lambda=K_1+K_2,%
\label{variable1}
\eeq
and
\beq
k_i=K_i-{1\over 2},\quad i=1,2,3.\label{variable2}
\eeq
The KP hierarchy is equivalent to (\ref{hbde1}) when the coefficients are
given by
\beq
\alpha=z_1(z_2-z_3)\quad\beta=z_2(z_3-z_1)\quad\gamma=z_3(z_1-z_2).
\label{coekp}
\eeq

We like to derive HBDE which is satisfied by Laughlin's functions. To
do this we rewrite it in the symmetric form, due to the conservation law
of momenta:
\beq
\tilde f(k_1,k_2,\ldots,k_N)=\prod_{i\ne j}(z_i-z_j)^{k_ik_j/2}\cdot%
  \exp\left[ i\pi\left(-{1\over 4}\sum_lk_l^2+{1\over 4}K^2\right)\right].
\label{veneziano2}
\eeq
The phase factor is nothing to do with the difference of $k_i$'s because of
the totally symmetric form. By substituting (\ref{veneziano2}) into
(\ref{hbde3}) directly we find
out that it is a solution to
HBDE provided the coefficients are given by
\barr
\alpha&=&2z_1-z_2-z_3\nonumber\\
\beta &=&2z_2-z_3-z_1\nonumber\\
\gamma&=&2z_3-z_1-z_2.\label{coel}
\earr
Since (\ref{veneziano2}) is symmetric under the exchange of variables the
same form of HBDE
holds for any three variables out of all $k_i$s. This means that the operator
(\ref{vo2}) is the generator of the B\"acklund transformation in the sense
that it
generates a new solution of HBDE with new set of variables we added.

Once we find the equation which characterizes Laughlin's wave function there
naturally arises a question, if there exist other solutions. To answer this
question we recall that HBDE with generic values of the coefficients have
been known to have soliton solutions. For instance the one soliton solution
takes the form
\beq
\tilde f(k_1,k_2,\ldots,k_N)=1+\exp\left(\sum_{i=1}^Np_ik_i+\varphi\right)%
\label{soliton1}
\eeq
The velocities $p_i$'s of this soliton are subject to the following spectral
condition
\beq
F(v_1,v_2,v_3)=0,\label{cond2}
\eeq
where
\barr
v_1&=&{1\over 2}\left(-p_1+p_2+p_3\right)\nonumber\\
v_2&=&{1\over 2}\left(-p_2+p_3+p_1\right)\nonumber\\
v_3&=&{1\over 2}\left(-p_3+p_1+p_2\right),\label{velocity}
\earr
which relates $p_i$'s to $z_i$'s through (\ref{variable1}) (\ref{variable2})
and (\ref{cond1}).
The explicit form of the spectral condition is the following,
\barr
(2z_1-z_2-z_3)\cosh\left[{1\over 2}(-p_1+p_2+p_3)\right]&&\nonumber\\
(2z_2-z_3-z_1)\cosh\left[{1\over 2}(-p_2+p_3+p_1)\right]&&\nonumber\\
(2z_3-z_1-z_2)\cosh\left[{1\over 2}(-p_3+p_1+p_2)\right]&=&0.%
\label{spectral}
\earr
In the form of soliton solution (\ref{soliton1}) $k_i$'s and $p_i$'s stand
for the coordinates and the velocities, respectively. As is shown $p_i$'s
depend on $z_i$'s via (\ref{spectral}). That is to say, the soliton has
a certain information of the QHE through the values of the filling factor
and the positions of electrons. The physical interpretation of their
correspondence is under consideration.
\vskip 1cm

In summary we like to remark the following:

\vskip 0.5cm
1) We have shown that  Laughlin's
wave function is possible to be
obtained by means of the solitonlike vertex operator (\ref{vo2}).
It allows, however, the soliton picture of \cite{Kashiwara}\nocite{Date1}-%
\cite{Date2} only at $k_i=1$.

2) We have shown that Laughlin's wave function is a solution to HBDE with
the coefficients of (\ref{coel}). This means that the operation of our vertex
operator (\ref{vo2})
causes the B\"acklund transformation. Namely, by performing the
operation one gets another solution.

3) As is known, HBDE of the KP hierarchy has a vast
solution space. It is natural that HBDE with the coefficients (\ref{coel}) has
also other solutions. For example we presented soliton solutions explicitly
which may represent suggestive properties of the QHE which may not be observed
up to now.

4)The string soliton correspondence was accomplished \cite{Saito}\cite{Sogo}
via the Miwa
transformation \cite{Miwa} which relates the momenta of strings to the time
variables in the soliton theories:
\beq
t_n={1\over n}\sum_jk_jz_j^n.\label{Miwat}
\eeq
The Laughlin theory is realized by restricting the values of $k_j$ to
$k_j^2=m$. We like to recall that the same type of restriction of $k_j$ in
(\ref{Miwat}) provides a basis of the 2D minimal conformal field theory
associated with the matrix model of the 2D gravity \cite{Chau}%
\nocite{Kon}-\cite{Semikhatov}.

The authors thank Prof. Han-Ying Guo for discussions.

\end{document}